\documentclass[
    ,final            
  ]
  {aipproc}

\layoutstyle{8x11single}

\begin{document}

\title{Other Exotic Scenarios at the LHC}

\classification{12.60.Cn, 12.60.Fr, 12.60.Rc, 12.90.+b, 14.80.Cp}
\keywords      {Exotics, Beyond the Standard Model}

\author{Kamal Benslama on behalf of the ATLAS Collaboration}{
  address={Physics Department, University of Regina, Saskatchewan, Canada}
}



\begin{abstract}
 The considerable center-of-mass energy and luminosity provided by the Large Hadron Collider (LHC) will ensure a discovery reach for new particles which extends well into the multi-TeV region. ATLAS and CMS have carried out many studies of the implications of this capability for Beyond the Standard Model physics. In this paper, we summarize some key results of studies involving non-susy models, such as extra-dimensions, little higgs, compositeness, and left-right symmetric models. 
\end{abstract}

\maketitle


\section{Introduction}

 The ATLAS \cite{atlas} and CMS \cite{cms} experiments at the LHC will give access to an unprecedented range of processes at $p_{T}$ scales and event rates far in excess of those generated by earlier colliders. Many rare Standard Model (SM) processes will be studied with high statistics, while at the same time searches will be carried out for particles predicted to exist by theories extending the SM. The need to stabilise the mechanism for breaking electroweak (EW) symmetry in the SM requires that at least some of these new particles have masses of the order of the EW scale ($\sim$ 100 GeV). Hence, there is every reason to believe that they should be observable at the LHC.

In this paper we review the results of some of the studies of Beyond the Standard Model physics at the LHC which have been carried out by ATLAS and CMS, concentrating on those involving extra dimensions, compositeness, little higgs, and lef-right symmetric models. For reasons of space, we do not discuss results of studies of technicolor models, but further details of this and other work can be found in Refs. \cite{techni} \cite{other1} \cite{other2}

\section{Extra Dimensions}

Models incorporating extra space-time dimensions (ED) have recently become extremely popular with phenomenologits. An attractive feature of many of these models is that the effective Plank Scale is brought down to the EW scale by power-law running of the gauge coupling constants, thereby removing the gauge hierarchy problem. A by-product of this running is that we expect quantum gravity effects to begin to mainifest themselves at the TeV scale, possibly within the range accessible to the LHC. Striking potential signatures can include graviton resonances contributing to Drell-Yan processes and new scalar bosons. The precise characteristics of these signatures depend strongly on the size, shape, and number of extra dimensions.

ATLAS and CMS have investigated three different classes of ED models incorporating Large \cite{largeED}, Warped \cite{warpED} or $TeV^{-1}$ scale \cite{smallED} extra dimensions. Large ($ >> TeV^{-1})$ extra dimension models such as that proposed by Arkani-Hamed, Dimopoulos and Dvali \cite{largeED}, typically assume the existence of $\delta$ EDs of size R in which only gravity can propagate. The observed Plank scale $M_{pl}$ following compactification is then related to the fundamental scale of gravity $M_{D}$ by the relation

\begin{equation}
M_{Pl}^{2} \; \sim \; R^{\delta} M_{D}^{2+\delta}
\end{equation}

If $M_{D}$ is near the EW scale then the gauge hierarchy problem is neatly solved. Such models predict the existence of an infinite number of Kaluza-Klein (KK) excitations of the graviton of very small mass splitting propagating in the $\delta$ dimensional bulk space-time. This leads to possible discovery signatures based on the searches for KK gravitons + quark (monojet) or photon production, in which the graviton is emmitted into the bulk leaving large amount of $E_{T}^{miss}$. An ATLAS study of the monojet signature \cite{laurent} has shown that after one year of high luminosity running, ATLAS should be able to confirm ADD models with $M_{D} < 9.1$, 7.0 or 6.0 TeV for 2, 3 or 4 EDs respectively. An alternative approach searching for associated production of KK gravitons with dileptons and diphotons \cite{kabachenko} has given equivalent reaches of 8.1, 7.9 and 7.4 TeV, indicating that potential signals may be observable in more than one channel.

Other ED models construct the universe from the usual four dimensions together with large EDs and a number of smaller $TeV^{-1}$ scale EDs \cite{smallED}. The SM fermions propagate only in the usual 4D space-time while the SM gauge bosons propagate additionally in the $TeV^{-1}$ scale EDs and gravitons propagate throughout. The dynamics associated with the $TeV^{-1}$ scale EDs leads to 4D KK excitations of the SM bosons. With 100$fb^{-1}$, peaks in dileptons invariant mass spectra (Fig. 1) arising from excitations of the first KK modes of the photon and $Z^{0}$ can be detected for compactification scale $M_{c} < 5.8 TeV$ and excesses above SM Drell-Yan tails for $M_{c} < 9.5 TeV$ \cite{giacomo}.

A third class of ED studies by ATLAS and CMS \cite{warp1} \cite{warp2} is the warped ED scenario of Randall and Sundrum \cite{warpED}. In these models the EW scale is generated from the Plank scale through warping of one small ED, rather than the presence of large flat EDs. The universe is here envisaged to consist of two 3-branes bounding a warped 5D bulk. The SM gauge fields propagate on one of the 3-branes while gravitons are localized on the other. Direct searches for new resonances in the dielectrons and diphotons mass spectra have been studied by CMS \cite{cmselec} \cite{cmsphot}, and it has been found that for a Randall-Sundrum model with coupling parameter c = 0.01, the graviton excitations have weak coupling to the Standard Model particles. Fig. 2 shows the CMS reach in the search for the Randall-Sundrum graviton decaying into diphoton (full line) and electron (dashed line) channels as a function of the coupling parameter c and the graviton mass for 10 $fb^{-1}$ and 30 $fb^{-1}$.

Another important finding by ATLAS \cite{warp1} is that these new states can be determined to be spin-2 rather than spin-1 for masses up to 1.7 TeV, allowing $Z^{\prime}$ interpretations of potential signals to be tested (Fig. 3).

\begin{figure}
  \includegraphics[height=.3\textheight]{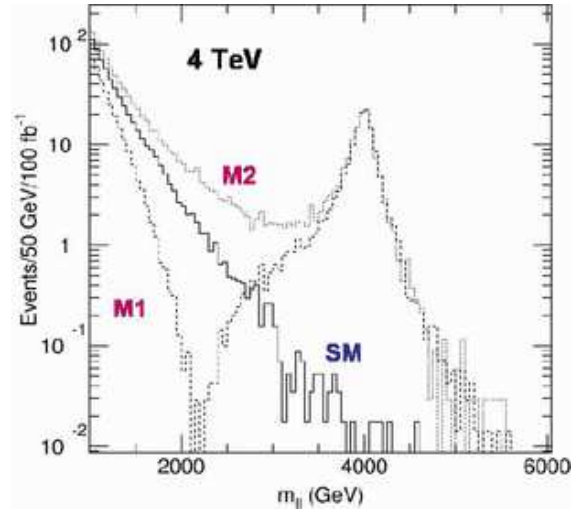}
  \caption{Invariant mass distribution of $e^{+}e^{-}$ pairs for the SM (full line) and for two models where $Z^{(1)}/\gamma^{(1)} \rightarrow e^{+}e^{-}$ and $m_{Z^{(1)}}$ = 4 TeV.}
\end{figure}

\begin{figure}
  \includegraphics[height=.25\textheight]{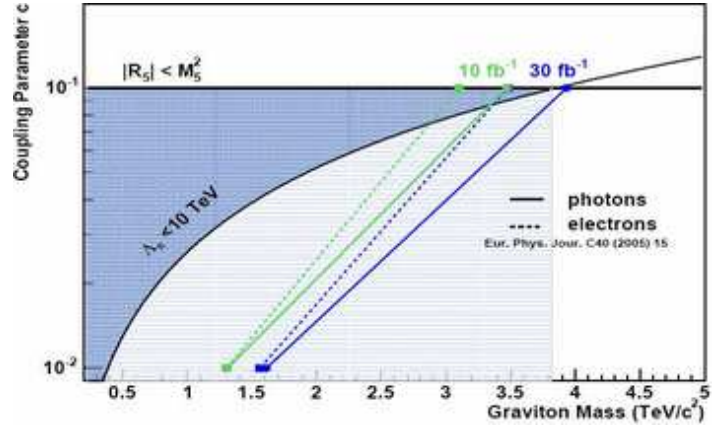}
  \caption{Reach of CMS experiment in the search for the Randall-Sundrum graviton decaying into diphoton and electron channels as a function of the coupling parameter c and the graviton mass for 10 $fb^{-1}$ and 30 $fb^{-1}$. The left part of each curve is the region where the significance exceeds 5 $\sigma$.}
\end{figure}

\begin{figure}
  \includegraphics[height=.3\textheight]{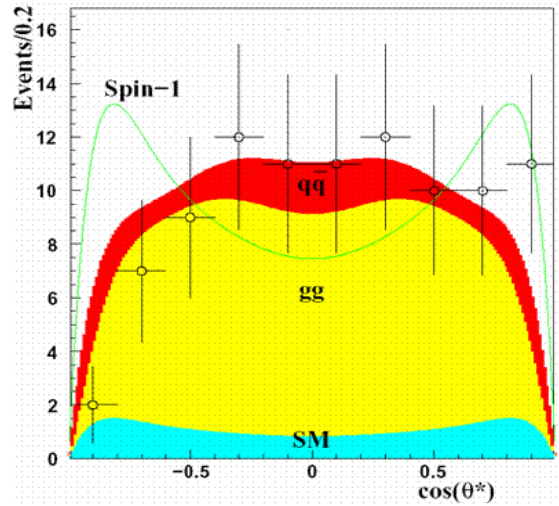}
  \caption{Angular distribution of $e^{+}e^{-}$ pairs for the graviton narrow resonance (open circle, yellow curve for the gg dominant production and red curve for the $q \bar{q}$ production), for the SM (bottom blue curve) and the expected distribution for a spin 1 resonance (green line).}
\end{figure}

\section{Excited quarks and leptons}

 A consequence of a possible substructure of quarks is the existence of excited quarks $q^{*}$ with masses of the order of the compositeness scale $\Lambda$.The excited quarks production at the LHC by quark-gluon fusion was studied for their subsequent decays to photon + jets \cite{excited1}, to quarks and gauge bosons and to quarks and gluons \cite{excited2} \cite{excited3}. For an integrated luminosity of $3 \times 10^{5} pb^{-1}$ and different coupling f determined by the compositeness dynamics, the $(q^{*} \rightarrow q W$) channel allows the discovery of new excited quarks up to masses of 7 TeV. For the ($q^{*} \rightarrow qZ)$ channel, the upper mass limit was found to be 4.5 TeV while for the ($q^{*} \rightarrow \gamma + jet)$ the upper mass limit was 6.5 TeV. For the channel ($q^{*} \rightarrow g q$), the upper mass limit was 6 TeV. Excited electrons \cite{excited4} can be accessible up to a mass of 3-4 TeV. Fig. 4 shows the invariant mass distribution of Ze ($Z \rightarrow jj$) for excited electrons masses of 500 GeV, 1 TeV, 2 TeV, 3 TeV and 4 TeV, for an integrated luminosity of 300 $fb^{-1}$ and $\Lambda = 6 \; TeV$.

\begin{figure}
  \includegraphics[height=.3\textheight]{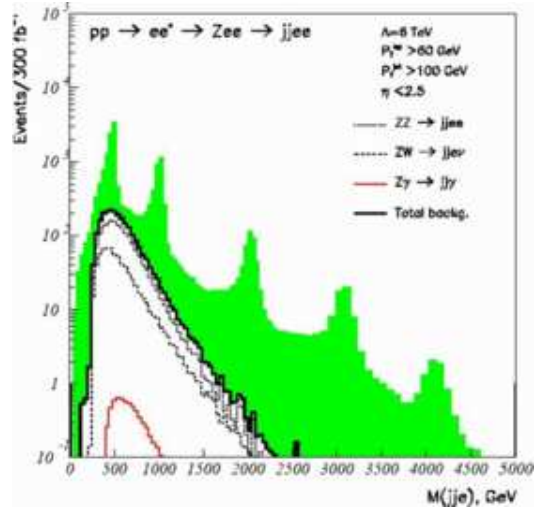}
  \caption{Invariant mass distributions of Ze ($Z \rightarrow jj)$ for excited lepton masses of 500 GeV, 1 TeV, 2 TeV, 3 TeV and 4 TeV, for an integrated luminosity of 300 $fb^{-1}$ and $\Lambda = 6 \;TeV$.}
\end{figure}

\section{Leptoquarks}

Among possible new particles in physics beyond the SM, leptoquarks (LQs) are an interesting category of exotic color triplets with coupling to quarks and leptons. They are a generic prediction of Grand Unified Theories \cite{lepto1}, of composite models \cite{lepto2}, of technicolor schemes \cite{lepto3}, of superstring-inspired $E_{6}$ models \cite{lepto4}, and of supersymmetry with R-parity violation \cite{lepto5}.

A recent ATLAS study \cite{lepto6} showed that the existence of TeV-scale leptoquarks can be probed for all fermion generations in pair production channels. In the $LQ\bar{LQ} \rightarrow l^{+}l^{-} q\bar{q}$ mode, the observation of a first or second generation leptoquark is feasible for up to $M_{LQ} \simeq 1.3$ TeV ($M_{LQ} \simeq 1 \;TeV$) with an integrated luminosity of 30 $fb^{-1}$, assuming $\beta =1 \;(\beta = 0.5)$ where $\beta$ is the leptoquark branching fraction defined by $\beta \equiv Br(LQ \rightarrow l q)$ and it is predicted by the model. Fig. 5 shows the $m_{ej}$ distributions for background (light line) and signal plus background (dark line), for various values of LQ mass and for an integrated luminosity of 30 $fb^{-1}$.

In the $LQ\bar{LQ} \rightarrow \nu_{\tau}\bar{\nu_{\tau}}b\bar{b}$ channel, on the other hand, third-generation leptoquarks coupling to a $\tau$-neutrino and b-quark (b anti-quark) with $\beta = 0.5 \;(\beta=0)$ will be equally probed for LQ masses of up to 1 TeV (1.3 TeV).

\begin{figure}
  \includegraphics[height=.3\textheight]{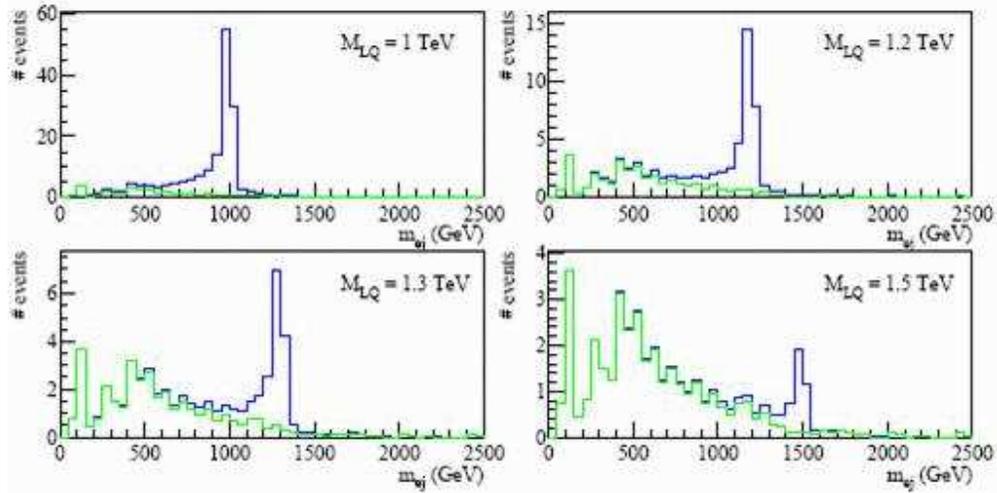}
  \caption{$m_{ej}$ distributions for background (light line) and signal plus background (dark line), for various values of LQ mass and for an integrated luminosity of 30 $fb^{-1}$.}
\end{figure}

\section{Little Higgs Models}

The Litte Higgs models implement the idea that the Higgs boson is a pseudo-Goldstone boson \cite{ps1} \cite{ps2}. They are constructed by embedding the SM inside a larger group with an enlarged symmetry. The larger symmetry is then broken at some high scale $\Lambda_{H} = 4 \pi f$, so that the Higgs mass is protected from radiative corrections at one loop that are dependent quadratically on $\Lambda_{H}$. From a phenomenological point of view, the effect of this symmetry is to require the existence of new particles whose coupling ensure that the large contributions to the Higgs mass are canceled. There must be three types of new particles (heavy top, new gauge bosons, and a higgs triplet which contains a doubly charged state) corresponding to the three contributions to radiative corrections to the Higgs mass. ATLAS studied \cite{little} the observability of these particles, and it has been found that the heavy top (T) can be observable up to masses of about 2.5 TeV via its decay to Wb. Sensitivity in Zt or ht is lower, but it still extends over the range expected in the model provided that the Higgs mass is not too large. The decay of T to Wb and Zt (fig. 6) are, of course, independent of the Higgs mass. In the case of ht, the sensitivity will depend on the Higgs mass.

In the case of the new gauge bosons, the situation is summarized in Fig. 7. The latter shows the accessible regions via the $e^{+}e^{-}$ final states of $Z_{H}$ and $A_{H}$ as a function of the mixing angles and the scale f that determines the masses. Except for a small region near $\tan\theta^{\prime}$ = 1.3 and $f > 7.5 TeV$ , we are sensitive to the whole range expected in the model. However, observation of such a gauge boson will not prove that it is of the type predicted in the Little Higgs Models. In order to do this, the decays to the SM higgs boson must be observed.

In the case of $\Phi^{++}$, the situation is not so promising. The Higgs sector is the least constrained by fine tuning arguments, and this particle's mass can extend up to 10 TeV. We are only sensitive to masses up to 2 TeV or so provided that $\nu^{\prime}$ is large enough. 

\begin{figure}
  \includegraphics[width=.4\textwidth]{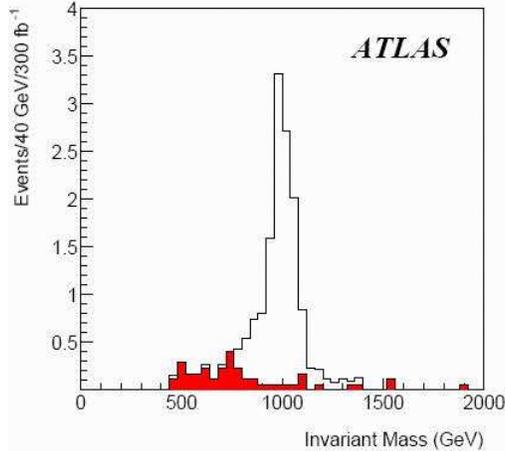}
  \caption{Reconstructed mass of the Z and t (inferred from the measured lepton, missing $E_{T}$, and tagged b-jet). The signal $T \rightarrow Z t$ is shwon for a mass of 1000 GeV. The background, shown as the filled histogram is dominated by WZ and tbZ production. The signal event rates corresponds to $\frac{\lambda_{1}}{\lambda_{2}} = 1$ and a $BR(T \rightarrow ht)$ of 25$\%$.}
\end{figure}

\begin{figure}
  \includegraphics[width=.35\textwidth]{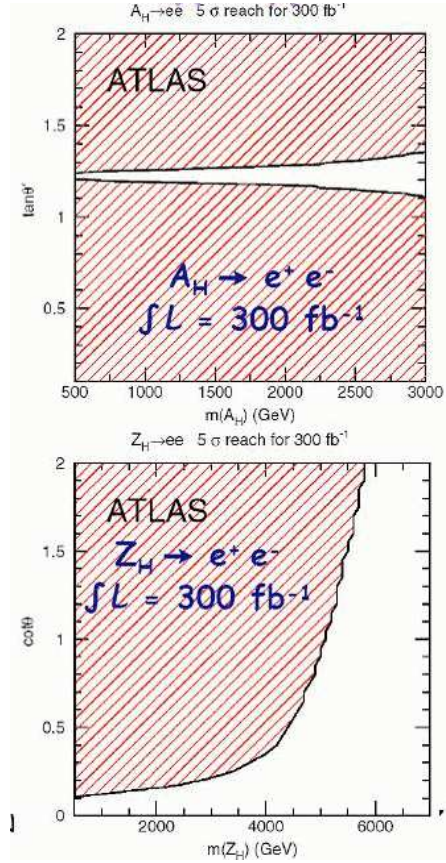}
  \caption{Upper plot shows the accessible region (shaded) in the channel $A_{H} \rightarrow e^{+}e^{-}$ as a function of the mass and the mixing $\tan\theta^{\prime}$. The lower plot shows the accessible region (shaded) in the channel $Z_{H} \rightarrow e^{+}e^{-}$ as a function of the mass and the mixing $\cot\theta$.}
\end{figure}

\section{Doubly Charged Higgs}

Doubly charged scalar particles arise in many scenarios \cite{hutu} extending the weak interaction beyond the Standard Model. In the left-right symmetric (LR) model \cite{lepto1} such particle is a member of a triplet Higgs representation which plays a crucial part in the model. The gauge symmetry $SU(2)_{L} \times SU(2)_{R} \times U(1)_{B-L}$ of the LR model is broken to the SM symmetry $SU(2)_{L} \times U(1)_{Y}$ due to a triplet Higgs $\Delta_{R}$, whose neutral component acquires a non-vanishing expectation value in the vacuum. The $\Delta_{R}$, called the right-triplet, consists of the complex fields $\Delta^{0}_{R}, \Delta^{+}_{R}$ and $\Delta^{++}_{R}$. If the langrangian is assumed to be invariant under a discrete $L \leftrightarrow R$ symmetry, it must contain, in addition to $\Delta_{R}$ also, a left-triplet $\Delta_{L} = (\Delta^{0}_{L}, \Delta^{+}_{L}, \Delta^{++}_{L})$. Hence, the LR model predicts two kinds of doubly charged particles with different interactions. In contrast with $\Delta_{R}$, the existence of $\Delta_{L}$ is not essential from the point of view of the spontaneous symmetry breaking of the gauge symmetry. The vacuum expectation value $v_{L}$ of its neutral member is actually quite tightly bound by the $\rho$ parameter, i.e. by the measured mass ratio of the ordinary weak bosons.

The main experimental signature of a produced $\Delta^{++}_{L,R}$ would be a hard same-sign lepton pair. There will be no substantial background due to the Standard Model, since in the SM a same-sign lepton pair will always be associated with missing energy, i.e. neutrinos, due to lepton number conservation. 

ATLAS has invetigated production of doubly charged Higgs particles $\Delta^{++}_{L,R}$ via the WW fusion process at the LHC in the framework of the left-right symmetric model \cite{doubly}. The production cross section of the right-triplet Higgs $\Delta^{++}_{R}$ is for representative values of model parameters at $fb$ level. The discovery reach depends on the mass of the right-handed gauge boson $W_{R}$. Fig 8 shows the discovery reach for $\Delta^{++}_{R} \rightarrow l^{+}l^{+}$ in the plane $m_{W^{+}_{R}}$ versus $m_{\Delta^{++}_{R}}$ (or $v_{R}$) for different luminosity scenarios, and assuming 100$\%$ BR to leptons. For $\Delta^{++}_{L}$, the discovery reach depends on the value of the left-triplet vev $v_{L}$ as shown on Fig. 9.

\begin{figure}
  \includegraphics[width=.4\textwidth]{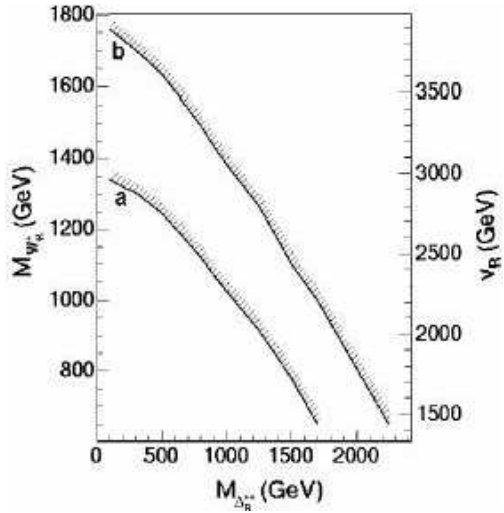}
  \caption{Discovery reach in the plane $m_{W_{R}}$ versus $m_{\Delta^{++}_{R}}$ (or $v_{R}$) in the decay $\Delta^{++}_{R} \rightarrow l^{+}l^{+}$ (l = e, $\mu$) for an integrated luminosity of 100 $fb^{-1}$ (a) and 300 $fb^{-1}$ (b), and assuming 100$\%$ BR to leptons. The region where the discovery is not possible is on the hatched side of the line.}
\end{figure}
\begin{figure}
  \includegraphics[width=.4\textwidth]{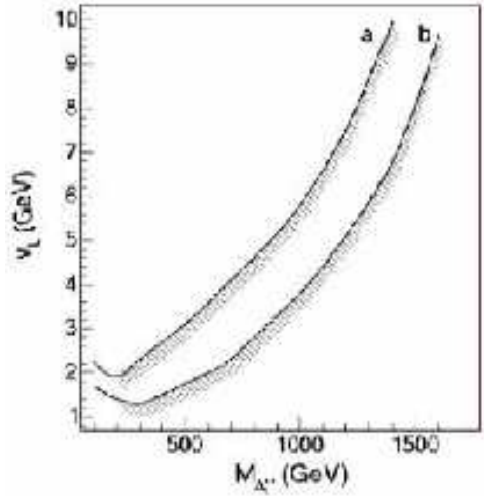}
  \caption{Discovery reach for $\Delta^{++}_{L} \rightarrow l^{+}l^{+}$ (l = e, $\mu$) in the plane $v_{L}$ versus $m_{\Delta^{++}_{L}}$ for an integrated luminosity of 100 $fb^{-1}$ (a) and 300 $fb^{-1}$ (b), and assuming 100$\%$ BR to leptons.}
\end{figure}

\begin{theacknowledgments}
 
 The author wishes to thank members of the ATLAS and CMS collaborations for the work described in this paper.
 
\end{theacknowledgments}


\end{document}